\documentstyle[11pt,newpaspIAU210,epsf]{article}

\newcommand{\HI}{H\,{\sc i}}

\newcommand{\SI}{S\,{\sc i}}
\newcommand{\SII}{S\,{\sc ii}}

\newcommand{\MgII}{Mg\,{\sc ii}}
\newcommand{\FeII}{Fe\,{\sc ii}}

\newcommand{\water}{\mbox{\rm H$_2$O}}
\newcommand{\feh} {\mbox{\rm [Fe/H]}}

\newcommand{\sfe} {\mbox{\rm [S/Fe]}}

\newcommand{\teff}  {\mbox{$T_{\rm eff}$}}
\newcommand{\kmprs}  {\mbox{\rm km\,s$^{-1}$}}

\newcommand{\logg}  {\mbox{{\rm log}\,$g$}}

\begin{document}
\title{Sulphur and iron abundances in halo stars\altaffilmark{1}}

\author{Poul E. Nissen}
\affil{Department of Physics and Astronomy, University of Aarhus, Denmark} 

\author{YuQin Chen}
\affil{National Astronomical Observatories, Beijing, China}

\author{Martin Asplund}
\affil{Mount Stromlo Observatory, Australian National University}

\author{Max Pettini}
\affil{Institute of Astronomy, Cambridge, UK}

% Notice that some of these authors have alternate affiliations, which
% are identified by the \altaffilmark after each name.  The actual alternate
% affiliation information is typeset in footnotes at the bottom of the
% first page, and the text itself is specified in \altaffiltext commands.
% There is a separate \altaffiltext for each alternate affiliation
% indicated above.

% The nice thing about this method is that it saves space on the first page!

% BUT if you've used \altaffiltext and you think you'll be adding 
% *footnotes*, then you need to update the footnote counter!

\altaffiltext{1}{Based on observations collected at the European Southern
Observatory, Chile (ESO No. 67.D-0106)}
\setcounter{footnote}{1}

% The abstract is entered in a LaTeX "environment", designated with paired
% \begin{abstract} -- \end{abstract} commands.  Other environments are
% identified by the name in the curly braces.

\begin{abstract}

From equivalent widths of the S\,{\sc i} lines at 8694\,\AA ,
Israelian \& Rebolo (2001) and Takada-Hidai et al. (2002)
have derived a surprisingly high
sulphur-to-iron ratio ([S/Fe] $\simeq$ 0.5 to 0.7) in six halo stars with
[Fe/H] $\simeq -2.0$ suggesting perhaps that hypernovae made a significant
contribution to the formation of elements in the early Galaxy.
To investigate this problem we have used
high-resolution spectra obtained with the ESO VLT/UVES
spectrograph to determine the S/Fe ratio in 19 main-sequence and
subgiant stars ranging in [Fe/H] from $-3.2$ to $-0.7$. The sulphur abundances
are determined from S\,{\sc i} lines at 8694\,\AA\  $\em and$
9212 - 9237\,\AA , and the iron abundances from about 20
Fe\,{\sc ii} lines.
S/Fe ratios as derived from 1D model atmospheres are presented and possible
3D effects are discussed.
The initial results from our survey do not confirm the high values
of [S/Fe] quoted above; instead we find that the ratio [S/Fe] remains
constant at about 0.35 dex for metallicities $-3 < \feh < -1$.

\end{abstract}

% Keywords should be included, but they are not printed in the hardcopy.
% They will be used by the Editors to help organize poster papers by
% category though!

\keywords{Stars: abundances, Stars: atmospheres, Galaxies: abundances}

% That's it for the front matter.  On to the main body of the paper.
% We'll only put in tutorial remarks at the beginning of each section
% so you can see entire sections together.

% 
% OK - to make things easier for the Editors, we're going to put
% all of our object aliases up front since we only have to declare
% them once in the paper.  Some people prefer to use NGC 7078 for
% M 15, but we like good old Messier, so that's what we'll index by.
% But we'll cross-reference it here so that people who do like NGC 7078
% won't have to remember that it's also M 15!
%
% Remember - we identify objects by putting an asterisk in front of the name!
%
%\index{*NGC 7078|M 15}

\section{Introduction}
Sulphur is generally regarded as an $\alpha$-capture element. The
work on Galactic stars by Fran\c{c}ois (1987, 1988)
supported this view by showing that [S/Fe]
increases from zero at solar metallicities to a plateau level
of about +0.5 dex in the metallicity range $-1.8 < \feh < -0.8$,
an analogous behaviour to that of other $\alpha$-elements
Mg, Si, and Ca, see Ryan et al. (1996). The
standard interpretation is that this trend arises from the
time delay in the production of 2/3 of the iron by SN of Type Ia
relative to the near-instantaneous release of the
$\alpha$-elements by Type II SN.

Recent observations of sulphur in metal-poor stars by
Israelian \& Rebolo (2001) have, however, challenged this view. Their data
suggest that \sfe\ increases linearly with decreasing \feh\ to a level
as high as $\sfe \sim +0.7$ at $\feh = -2.0$. The study of Takada-Hidai et al.
(2002) based on Keck HIRES observations 
supports a quasi linear dependence of \sfe\ on \feh\
although in their case \sfe\ reaches only 0.5 dex at $\feh = -2.0$.

As a possible explanation of the high value of \sfe\ in metal-poor stars,
Israelian \& Rebolo propose that massive supernovae with exploding
He-cores and a high explosion energy
make a significant contribution to the early chemical evolution
of galaxies. According to Nakamura et al. (2001) these hypernovae
overproduce S with respect to O, Mg and Fe. With
this intriguing possibility in mind a more thorough investigation
of sulphur abundances in halo stars seems worthwhile.
A clarification of the trend of S abundances is also much needed
in deciphering the chemical enrichment of
damped Ly$\alpha$ systems (DLAs), widely regarded as the
progenitors of present-day galaxies at high redshift. Its
importance stems from the fact that, unlike most other heavy
elements, S is not depleted onto dust. Consequently, observations
of the relatively weak \SII\ triplet resonance lines at $\lambda\lambda
1250, 1253, 1259$ yield a direct measurement of the abundance of
S in DLAs. The only other element for which this is the case is
Zn, and indeed most of our current knowledge of the chemical
evolution of the universe at high redshift is based on surveys of
[Zn/H] in DLAs (e.g. Pettini et al. 1999, Prochaska \& Wolfe 2002).
{\it If} S is an $\alpha$-capture
element, then its abundance relative to Zn (an iron-peak element)
could be used as `a chemical clock' to date the star-formation
process at high $z$. Specifically, if the major star formation
episodes in the DLAs observed occurred within the last $\approx
0.5$\,Gyr, we would expect to measure enhanced [S/Zn] ratios, and
{\it vice versa}.
Data on the abundance of S in DLAs have been relatively scarce
until recently, but are now becoming available at a progressively
faster rate thanks largely to the high sensitivity of the VLT/UVES
high resolution spectrograph at
blue and near-UV wavelengths. The picture is still confused
(Centuri\'{o}n et al. 2000, Prochaska \& Wolfe 2002),
but one thing is clear: without a secure
knowledge of the behaviour of S in metal-poor Galactic stars we
stand no hope of interpreting the situation at high $z$.

In this paper we report on a large scale survey
of sulphur abundances in metal-poor stars carried out with VLT/UVES.
Effects on the derived S/Fe ratio
from the modelling of stellar atmospheres are discussed, and
preliminary results for about half of the 35 stars observed
are presented.
 
\section{Observations and data reduction}
Previous determinations of sulphur abundances in metal-poor stars,
including those of Israelian \& Rebolo (2001) and Takada-Hidai et al. (2002),
have been based on the high excitation ($\chi = 7.87$\,eV)
$\lambda \lambda$8693.9, 8694.6 \SI\ lines. They are,
however, very weak in metal-poor dwarfs and giants ($W <$ 3\,m\AA\ at
$\feh \sim -2.0$), and practically impossible to observe at
$\feh \sim -2.5$. In our survey we have taken advantage of the
high efficiency of the VLT/UVES instrument and concentrated on
the stronger \SI\ triplet ($\chi = 6.52$\,eV) at 9212 - 9238\,\AA .
The dichroic mode of UVES was used to cover the spectral region
6700 - 10000\,\AA\ in the red arm and 4000 - 5000\,\AA\
in the blue arm; the latter includes a number of \FeII\
lines suitable for determining the iron abundance.
The resolution of the spectra is $\lambda/\Delta\lambda \simeq 60\,000$
with 4 pixels per spectral resolution element. Typical $S/N$ ratios are
300 in the blue region and 200 at the 9212 - 9238\,\AA\ triplet.

The red spectra were reduced using the echelle package in
IRAF. A problem with the 9212 - 9238\,\AA\ region is the presence
of numerous, quite strong telluric \water\ lines, see Fig.\,1.
To remove these lines fast rotating, early-type stars were observed on
each night and reduced in the same way as the program stars.
The IRAF task `$\em telluric$' was then used to remove the telluric lines.
When the early-type star is observed at about the same airmass as the program
star this technique also serves to remove a residual fringing
remaining from the flatfielding of the spectra.

\begin{figure}
\plotone{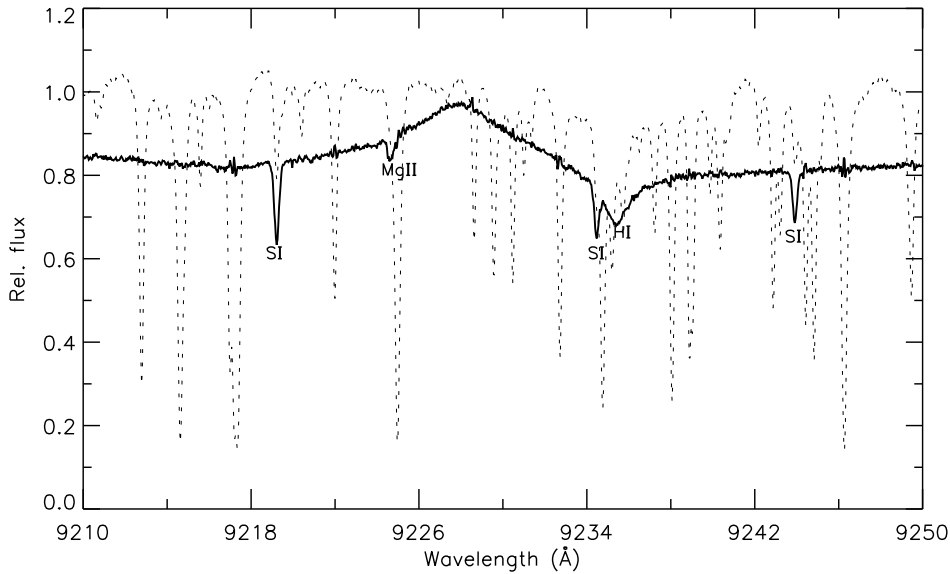} 
\caption{The VLT/UVES spectrum of HD\,110621 ($\feh = -1.66$).
The dotted line shows
the spectrum before removal of the numerous telluric \water\ lines. The thick,
full drawn line is the spectrum after division with the spectrum of
the B-type star HR\,5488 using the IRAF task `$\em telluric$' to obtain the
best fit between the two sets of \water\ lines. The radial velocity
of HD\,110621 was 207.1\,\kmprs , so all stellar lines are shifted
by about 6.4\,\AA . The \SI\ triplet, a \MgII\ line and the
Paschen $\zeta$ \HI\ line of HD\,110621 are seen. The broad
`emission' feature is due to the \HI\ line in the
B star spectrum. Excessive noise in the stellar spectrum are seen
where strong \water\ lines have been removed.}
\end{figure}

\begin{figure}
\plotone{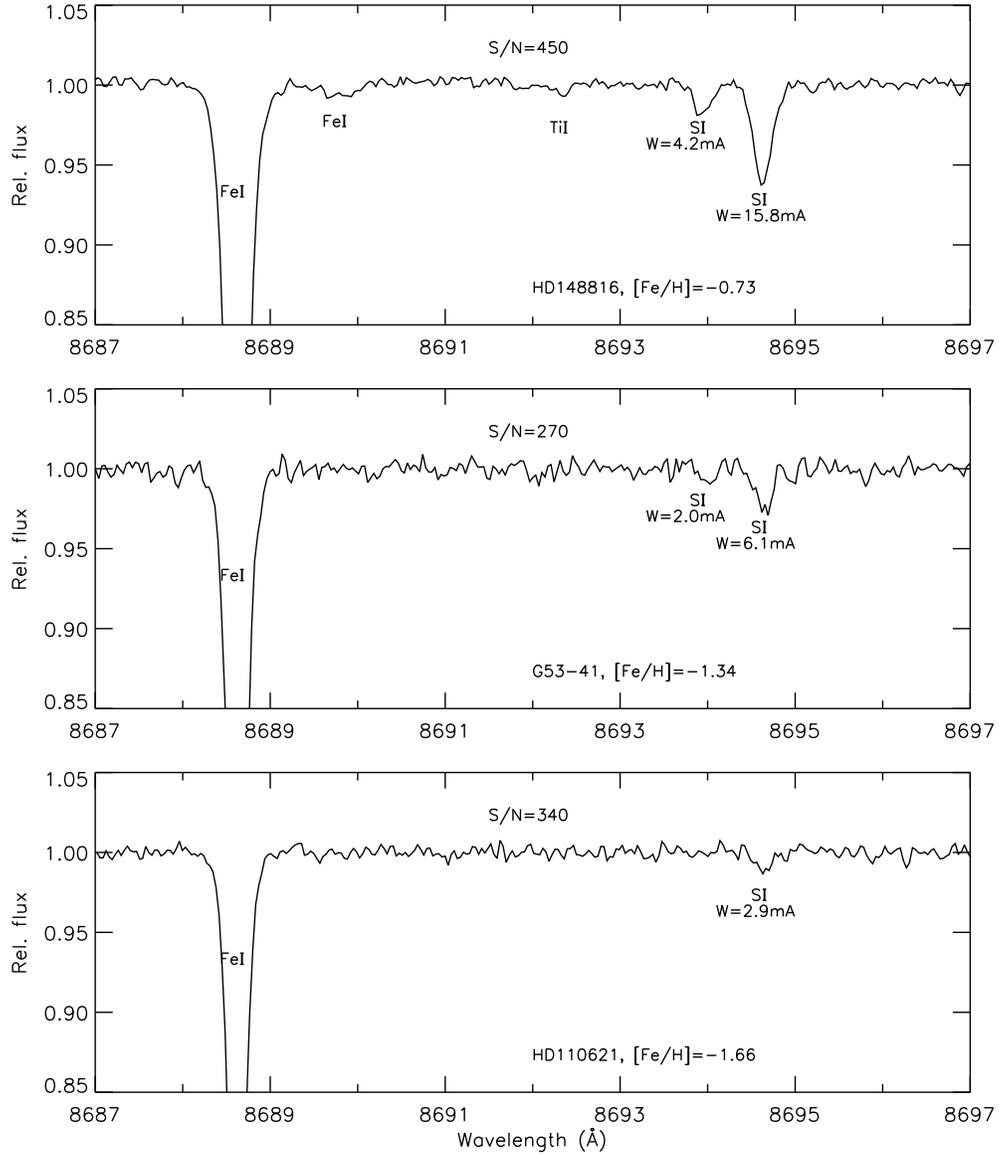}
\caption{A sequence of spectra around the
$\lambda \lambda$8693.9, 8694.6 \SI\ lines.}
\end{figure}

\begin{figure}
\plotone{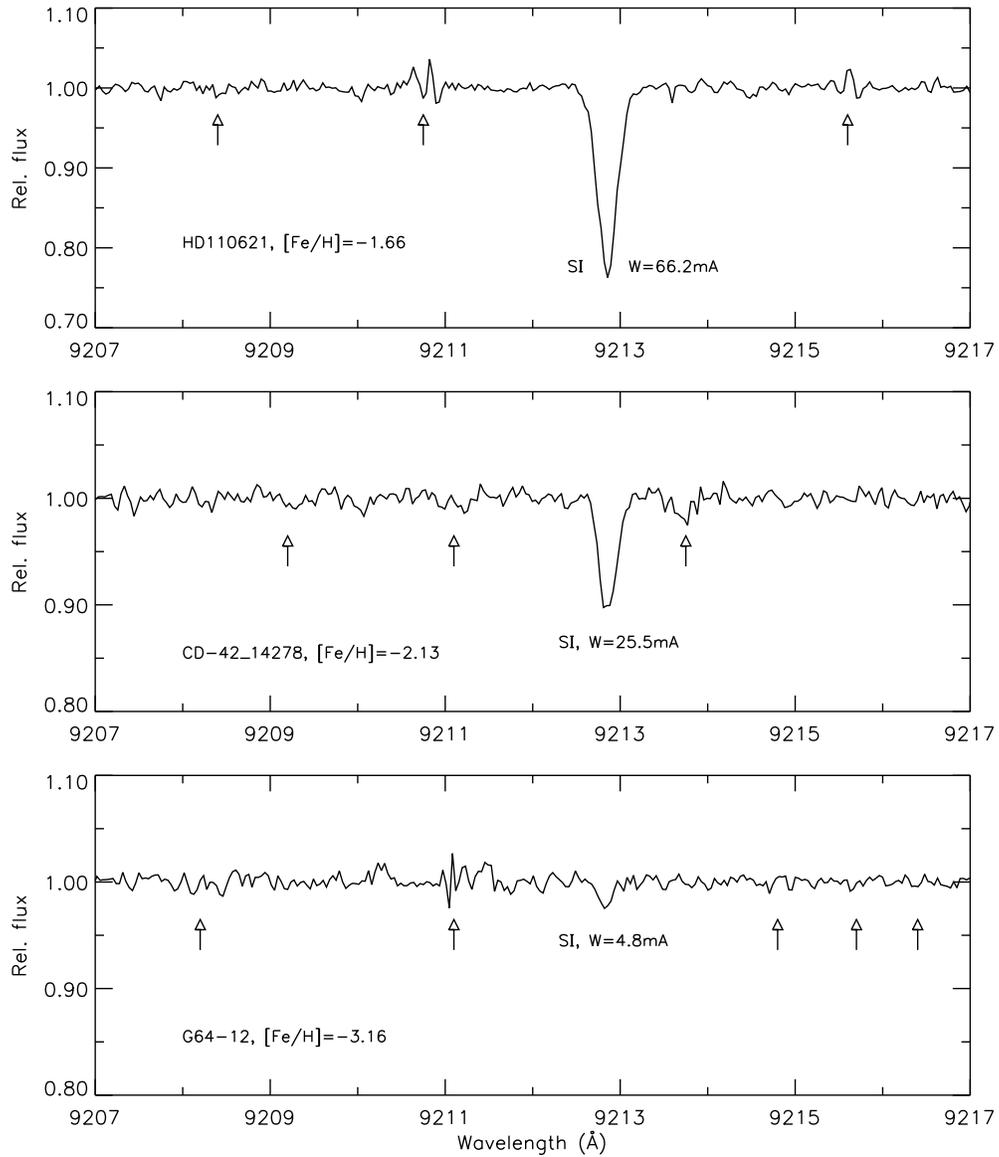}
\caption{ A sequence of spectra around the
$\lambda$9212.9 \SI\ line. The arrows indicate positions where telluric
\water\ lines have been removed. The spectra have $S/N$ around 200.}
\end{figure}

Representative spectra of stars are shown in Fig.\,2 (8694\,\AA\ region)
and Fig.\,3 (9212\,\AA\ region). As seen the \SI\ line at 9212.9\,\AA\
can be clearly detected in a $\feh \sim -3.2$ star. Eight stars were
observed twice on different nights; a comparison shows that the equivalent
widths of the $\lambda \lambda$8693.9, 8694.6 lines are measured with
an accuracy of about $\pm 0.5$\,m\AA\ whereas the $\lambda \lambda$9212.9,
9237.5 lines are measured to an accuracy of $\pm 2$\,m\AA .

The blue spectra were reduced with MIDAS routines. For a few stars it
was checked that IRAF reduction gives practically the same result.
These blue spectra have $S/N \sim 300$ and allowed us to measure
accurate equivalent widths of about 20 weak \FeII\ lines.

% In this section, we see the use of the \subsection command to set off
% an independent subsection.  We only have one here; usually there would
% be several.
%
% We show the use of several of the displayed math environments described
% in the User Guide, and you get a healthy dose of mathematical typesetting
% examples.  Also, observe the use of the LaTeX \label command after the
% \subsection to give a symbolic tag to the subsection for cross-referencing
% in a \ref command.  LaTeX automatically numbers the sections, equations,
% tables, etc. as it goes, so in general you don't know what number something
% is going to have.  We'll refer to the "hairymath" section a little later.

\section{Modelling of the stellar atmospheres}
The measured equivalent widths were used to derive S and Fe abundances
based on 1D MARCS (Asplund et al. 1997) model atmospheres assuming LTE.
Effective temperatures were derived from the color indices
$b-y$ and $V-K$, and surface gravities via Hipparcos parallaxes
and/or the Str\"{o}mgren $c_1$ index. For details we refer to the
recent paper by Nissen et al. (2002) on oxygen abundances.

Oscillator strengths of the blue \FeII\ lines were determined from an inverted
abundance analysis of three stars (HD\,103723, HD\,160617 and HD\,140283)
adopting \feh\ values determined by Nissen et al. (2002) from
very accurate equivalent widths of weak \FeII\ lines analyzed differentially
with respect to the solar spectrum. This ensures that we are on the same
\feh\ scale as Nissen et al. (2002).

Oscillator strengths of the sulphur lines were taken from Lambert \& Luck
(1978). We note that the values for the 8693.9 and 8694.6\,\AA\
lines (log$gf$ = $-0.56$ and 0.03, respectively) and the solar spectrum
equivalent widths provide a sulphur abundance of the Sun very close
to the meteoritic value, log$\epsilon$(S) = 7.20. The 9212.9 and
9237.5\,\AA\ \SI\ are too strong in the solar spectrum to provide
reliable abundances, but we note that the sulphur abundance derived
from this pair of lines agree well with that derived from the
$\lambda \lambda$8693.9, 8694.6 pair for 12 stars in the metallicity
range $-1.5 < \feh < -0.7$.
The mean difference between the S abundance derived from the
$\lambda \lambda$8693.9, 8694.6 pair and the 9212.9,9237.5\,\AA\ lines
is +0.03 dex with an rms deviation of the difference of $\pm 0.10$ dex.
The third line of the near-infrared \SI\ triplet at 9228.1\,\AA\
is unfortunately very close to the Paschen $\zeta$ \HI\ line at
9229.0\,\AA\ and could not be used for a reliable determination of
the sulphur abundance.

Errors in \teff\ ($\pm 70$\,K) and \logg\ ($\pm 0.15$ dex) have only
a minor effect on the derived \sfe\ (less than 0.05 dex). This is connected
to the fact that the abundances of both elements are determined from their
main ionization stages, \SI\ and \FeII ,  respectively.
The microturbulence was determined from requiring that the abundance
derived from the 20 \FeII\ lines should be independent of $W$, or
in the case of stars with very weak \FeII\ lines ($\feh < -1.5$)
$\xi_{\rm micro}$ is assumed to be 1.5\,\kmprs . In any case, the effect of the
uncertainty of $\xi_{\rm micro}$ on the S abundance is quite negligible
due to the weakness of the \SI\ lines.

As shown by Takada-Hidai et al. (2002) non-LTE effects on the
weak, high excitation \SI\ lines are likely to be small ($< 0.05$\,dex).
The same is the case for the \FeII\ lines according to Th\'evenin \& Idiart
(1999).

\begin{table}
\caption{1D - 3D model atmosphere effects on sulphur abundances. 
For each set of model atmosphere parameters the table first gives
the assumed 1D sulphur abundances ($A$ = log$\epsilon$(S)), then the 
computed 1D equivalent widths for the three most important
\SI\ lines and the 3D abundances derived from these equivalent widths.}
\begin{center}\footnotesize
\begin{tabular}{ccrcrcrcrc}
\tableline
        &       &      &          & 8694.6\,\AA &      & 9212.8\,\AA &
    & 9237.5\,\AA  &       \\
 \teff  & \logg & \feh & $A_{1D}$ & $W_{1D}$(m\AA ) & $A_{3D}$ & 
 $W_{1D}$(m\AA )& $A_{3D}$ & $W_{1D}$(m\AA )& $A_{3D}$  \\
\tableline
       &      &         &       &      &      &       &      &       &      \\
  5767 & 4.44 &    0.0 &  7.33 & 35.7 & 7.29 & 143.5 & 7.32 & 109.3 & 7.32 \\
  5822 & 4.44 & $-$1.0 &  6.73 & 15.2 & 6.75 & 112.4 & 6.75 &  81.2 & 6.75 \\
  5837 & 4.44 & $-$2.0 &  5.73 &  1.7 & 5.79 &  42.5 & 5.86 &  23.3 & 5.84 \\
  5890 & 4.44 & $-$3.0 &  4.73 &  0.2 & 4.78 &   7.1 & 4.81 &   3.1 & 4.81 \\
       &      &         &       &      &      &       &      &       &      \\
  6191 & 4.04 &    0.0 &  7.33 & 55.6 & 7.36 & 160.9 & 7.44 & 127.3 & 7.42 \\
  6180 & 4.04 & $-$1.0 &  6.73 & 25.8 & 6.77 & 114.0 & 6.74 &  88.1 & 6.76 \\
  6178 & 4.04 & $-$2.0 &  5.73 &  3.5 & 5.77 &  51.9 & 5.82 &  32.7 & 5.82 \\
  6205 & 4.04 & $-$3.0 &  4.73 &  0.4 & 4.77 &  11.4 & 4.81 &   5.2 & 4.80 \\ 
\tableline
\end{tabular}
\end{center}
\end{table}

Much attention has recently been paid to the effect on abundance
determinations of 3D hydrodynamical modelling of convection and granulation
in stellar atmospheres. As first shown by Asplund et al. (1999) there
is a very significant surface cooling in metal-poor stars caused by the
adiabatic cooling of rising and expanding elements coupled with the lack of
radiative heating due to the low line opacity in metal-poor stars.
The effect on lines formed in the upper atmosphere can be very large.
As an example Asplund \& Garc\'{\i}a P{\'e}rez (2001) estimated 3D effects
on oxygen abundances derived from OH lines of the order of $-0.6$\,dex
for turnoff stars with $\feh < -2.0$.

To investigate the 3D effects on the sulphur lines used in the present
study a differential 1D - 3D study similar to that described by
Asplund \& Garc\'{\i}a P{\'e}rez (2001) has been carried out. The
results are given in Table 1 for two sets of model atmospheres with
metallicities of $\feh = 0.0, -1.0, -2.0$ and $-3.0$. The first set has
\teff\ and \logg\ close to the solar values, whereas the second set
represents stars close to the turnoff. For each model atmosphere
Table 1 first gives the adopted S abundance, then the corresponding
computed 1D equivalent width, and finally the S abundance calculated
from the 1D equivalent width
on the basis of the 3D model atmosphere. As seen the maximum 3D
effect occurs for the most metal-poor stars: the 3D abundance is increased
by 0.07 to 0.13 dex with respect to the 1D abundance in the case of the
$\lambda \lambda$9212.9, 9237.5 \SI\ lines. Interestingly, there is, however,
a similar 3D effect ($\sim +0.10$\,dex) on the Fe abundance derived from
\FeII\ lines (Nissen et al. 2002). Hence, it appears that we are in
the favourable situation that the S/Fe ratio is quite immune to
1D - 3D effects.

\section{Results}
Up to now we have reduced and analyzed about half of the VLT/UVES spectra.
The derived sulphur abundances for 19 halo stars
are given in Table 2 together with
atmospheric parameters and iron abundances. Fig.\,4 shows a plot of
[S/Fe] vs. [Fe/H] with disk stars from Chen et al. (2002) added.
The sulphur abundances of the disk stars were derived from
the $\lambda \lambda$8693.9, 8694.6 \SI\ lines and three other weak
\SI\ lines in the spectral range 6000 -- 6800\,\AA .

\begin{table}
\caption{Effective temperatures, surface gravities, iron
    and sulphur abundances for 19 stars with halo kinematics.}
\begin{center}\footnotesize
\begin{tabular}{lccccrc}
\tableline
     &        &       &      &       &      &      \\
Star & \teff  & \logg & \feh & [S/H] & \sfe & Note \\
\tableline
                &     &     &        &        &        & \\
    HD103723    &6034 &4.26 &$-$0.81 &$-$0.84 &$-$0.03 & $\alpha$$-$def. \\
    HD105004    &5919 &4.36 &$-$0.86 &$-$0.85 & 0.01 & $\alpha$$-$def. \\
    HD106038    &5919 &4.30 &$-$1.42 &$-$1.00 & 0.42 & \\
    HD108177    &6039 &4.25 &$-$1.74 &$-$1.39 & 0.35 & \\
    HD110621    &5981 &3.99 &$-$1.66 &$-$1.34 & 0.32 & \\
    HD121004    &5636 &4.33 &$-$0.76 &$-$0.45 & 0.31 & \\
    HD140283    &5692 &3.69 &$-$2.41 &$-$2.11 & 0.30 & \\
    HD146296    &5666 &4.18 &$-$0.72 &$-$0.54 & 0.18 & \\
    HD148816    &5828 &4.14 &$-$0.73 &$-$0.50 & 0.23 & \\
    HD160617    &5931 &3.77 &$-$1.80 &$-$1.40 & 0.40 & \\
    G11-44      &5994 &4.29 &$-$2.08 &$-$1.71 & 0.37 & \\
    G13-09      &6361 &4.00 &$-$2.27 &$-$1.81 & 0.46 & \\
    G16-13      &5603 &4.17 &$-$0.75 &$-$0.47 & 0.28 & \\
    G20-08      &5855 &4.16 &$-$2.27 &$-$1.86 & 0.41 & \\
    G53-41      &5829 &4.16 &$-$1.34 &$-$1.00 & 0.34 & \\
    G64-12      &6512 &4.39 &$-$3.16 &$-$2.80 & 0.36 & \\
    G66-30      &6345 &4.24 &$-$1.52 &$-$1.30 & 0.22 & \\
 BD$-$13\,3442  &6501 &4.18 &$-$2.59 &$-$2.30 & 0.29 & \\
 CD$-$42\,14278 &5804 &4.25 &$-$2.13 &$-$1.82 & 0.31 & \\
\tableline
\end{tabular}
\end{center}
\end{table}

As seen from Fig.\,4 \sfe\ shows a clear plateau ($\sfe \simeq 0.35$)
for \feh\ below $-1.2$. There is no tendency for a linear rise of
\sfe\ as claimed by Israelian \& Rebolo (2001) and Takada-Hidai et al.
(2002). We note in this connection that their samples include
only three stars with $\feh < -2.0$ and none below $\feh = -2.3$, whereas our
sample includes seven stars in the range $-3.2 < \feh < -2.0$. Furthermore,
the error bars on their sulphur abundance determinations are large
due to the weakness of the \SI\ 8694.6\,\AA\ line; the 9212.8 and
9237.5\,\AA\ \SI\ lines, on the other hand, are rather ideal for
sulphur abundance determinations below $\feh = -2.0$. Although
we cannot exclude the existence of a few stars with high values of 
S/Fe\footnote{We do confirm the high value $\sfe \simeq 0.7$ of HD\,19445 
($\feh \simeq -1.9$) if we adopt the equivalent width (2.7\,m\AA )
of the \SI\ 8694.6\,\AA\ line given by Israelian \& Rebolo based on their
very high S/N spectrum} it would appear that 
sulphur shows the same trend as other typical
$\alpha$-capture elements Mg, Si and Ca. Hence, our preliminary conclusion
is that there is no need to invoke element production by hypernovae or
very massive supernovae to explain the general behaviour of sulphur. Traditional
Galactic evolution models with near-instantaneous production
of $\alpha$-elements
by Type II supernovae and delayed production of the iron-peak
elements are in good agreement with our data for sulphur (Chiappini et al.
1999, Goswami \& Prantzos 2000).

\begin{figure}
\plotone{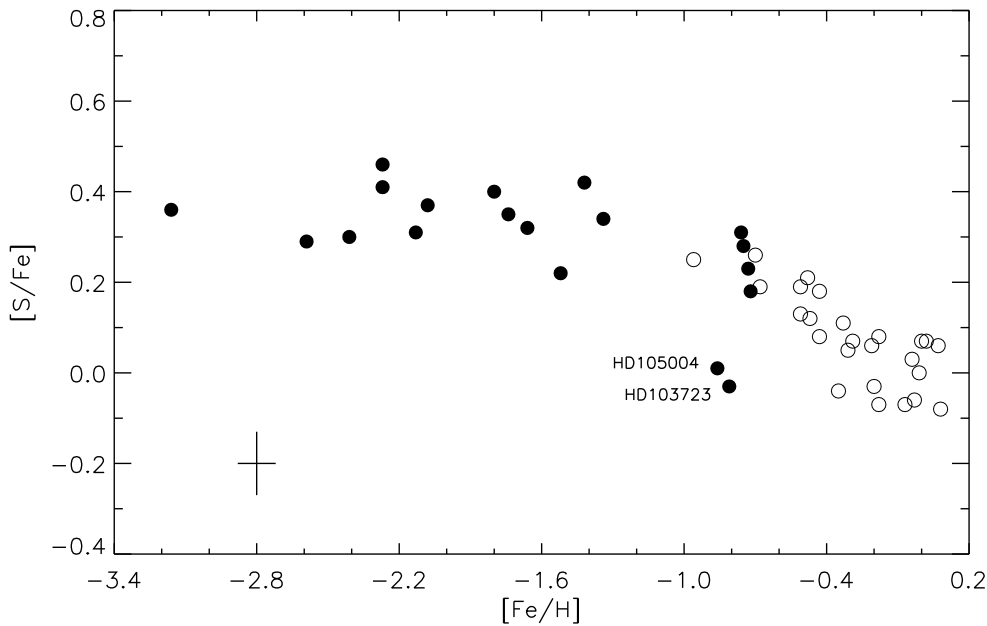}
\caption{\sfe\ vs. \feh . Filled circles are halo stars from the present
work and open circles are disk stars from Chen et al. (2002).
The typical error bars on the abundance ratios are shown, and
the two $\alpha$-deficient halo stars, HD\,103723 and HD\,105004
(Nissen \& Schuster 1997), are indicated.}.
\end{figure}

An interesting detail in Fig.\,4 should be noted. All stars from the
present investigation (plotted with filled circles) have halo kinematics,
including six stars with $-0.9 < \feh < -0.7$.
They have Galactic rotational velocities less than 50\,\kmprs\, i.e.
well below the characteristic rotational velocity of disk stars
(225\,\kmprs ) and thick disk stars (175\,\kmprs ). As seen from the figure,
there is then an overlap between halo stars and thick disk stars in
the metallicity range $-1.0 < \feh < -0.6$. Out of the six stars four
have enhanced S/Fe ratios like the thick disk stars but two
(HD\,103723 and HD\,105004) have a solar S/Fe ratio. Interestingly,
HD\,103723 and HD\,105004 were also found to be deficient in other
$\alpha$-element (O, Mg, Si, Ca, Ti) by Nissen \& Schuster (1997).
On the basis of their Galactic orbits, these authors speculated that the
$\alpha$-deficient stars have been accreted from dwarf galaxies
with a slow chemical evolution in contrast to the `normal' halo stars
with enhanced [$\alpha$/Fe] formed in a fast chemical evolution
in the inner part of our Galaxy. The fact that S and the
classical $\alpha$-elements, Mg, Si and Ca, are deficient in the same
stars is a further indication that sulphur belongs to this group of
elements. Thus there seems to be little ground, on the basis
of our data, for the reservations expressed by
Prochaska et al. (2000) concerning
use of S as an $\alpha$-element in the analysis
of abundance ratios in DLAs. On the other hand, one should remain
aware that the overabundance of the $\alpha$-elements
at low metallicities need not be a universal phenomenon.
The finding by Nissen \& Schuster (1997)
that even some stars in the Milky Way do not fit this pattern
is a warning against the naive assumption that such an overabundance
should also apply to all metal-poor DLAs.

\end{document}